\documentclass[12pt]{article}

\usepackage{scicite}

\usepackage{times}
\usepackage{hyperref}

\usepackage{color}

\definecolor{Gray}{gray}{.25}

\usepackage{graphicx}

\usepackage{sidecap}

\usepackage{textgreek}
\usepackage{siunitx}
\DeclareSIUnit\molar{\textsc{M}}

\usepackage{nicefrac} 
\usepackage{gensymb}

\usepackage{amsmath}

\usepackage{wrapfig}
\usepackage[pscoord]{eso-pic}
\usepackage[fulladjust]{marginnote}
\reversemarginpar

\newcounter{lastnote}

\title{Biological computations: limitations of attractor-based formalisms and the need for transients }

\begin{document} 

\maketitle

\author{Daniel Koch$^{1^\dagger}$, Akhilesh Nandan$^{1^\dagger}$, Gayathri Ramesan$^{1^\dagger}$ and Aneta Koseska$^{1,^\ast}$} \\
\\
\normalsize{$^{1}$  Lise Meitner Group Cellular Computations and Learning,\\ Max Planck Institute for Neurobiology of Behaviour - caesar, Bonn, Germany}\\
\\
\normalsize{$^\dagger$These authors contributed equally.}\\
\normalsize{$^\ast$To whom correspondence should be addressed; E-mail:  aneta.koseska@mpinb.mpg.de}


\date{}

\baselineskip24pt




\section*{Abstract}

Living systems, from single cells to higher vertebrates, receive a continuous stream of non-stationary inputs that they sense, e.g., via cell surface receptors or sensory organs. Integrating these time-varying, multi-sensory, and often noisy information with memory using complex molecular or neuronal networks, they generate a variety of responses beyond simple stimulus-response association, including avoidance behavior, life-long-learning or social interactions. In a broad sense, these processes can be understood as a type of biological computation. Taking as a basis generic features of biological computations, such as real-time responsiveness or robustness and flexibility of the computation, we highlight the limitations of the current attractor-based framework for understanding computations in biological systems. We argue that frameworks based on transient dynamics away from attractors are better suited for the description of computations performed by neuronal and signaling networks. In particular, we discuss how quasi-stable transient dynamics from ghost states that emerge at criticality have a promising potential for developing an integrated framework of computations, that can help us understand how living system actively process information and learn from their continuously changing environment.

\section*{Introduction}

\begin{figure}[h]
\includegraphics[width=\textwidth]{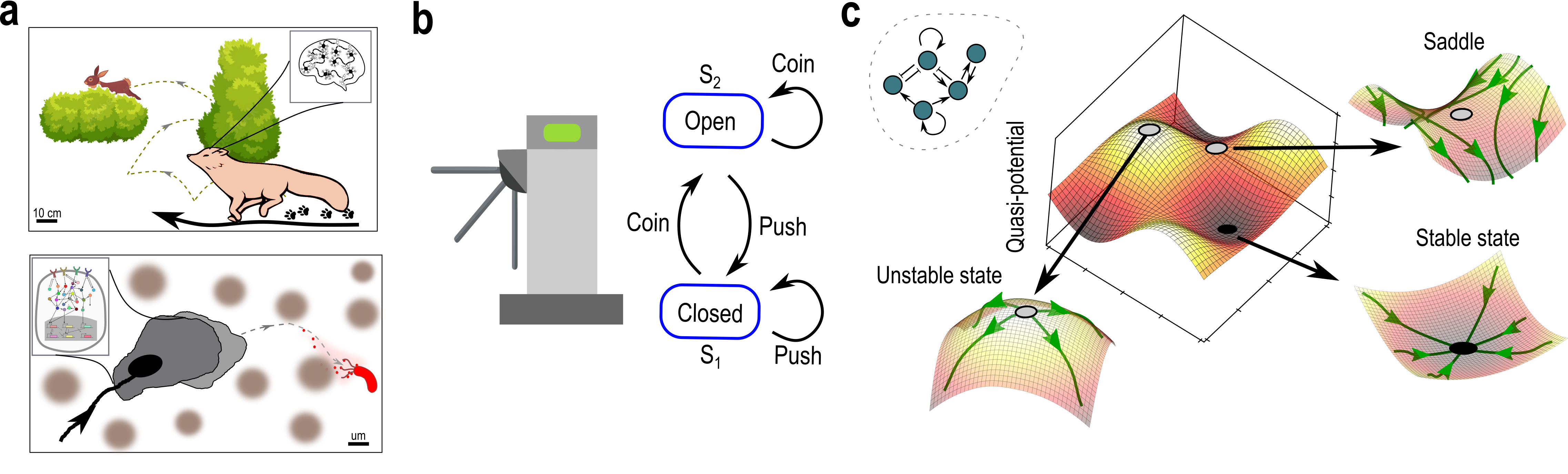}
\caption{Examples of natural  and machine computations. (a) Top: A predator hunting its prey integrates time-varying (for e.g. the prey escapes and/or hides) and multi-sensory (visual, olfactory, auditory etc.) inputs and continually adapts its response in real-time. A large fraction of the animal kingdom relies in the brain's neuronal network to process such information.
Bottom: A leukocyte chasing a bacterium faces similar challenges, yet relies on networks comprising entities of a completely different nature (i.e. molecules such as proteins, nucleic acids etc) to solve this task. Despite large differences in scale, type and phylogeny, the similarity of the tasks from a information processing perspective implies similar dynamical principles to underline the computational process. Gray circles: red blood cells. Dashed/solid lines: Prey(bacteria)/predator(leukocyte) tracks.  (b) Turnstile, an example of a finite-state-machine, and a schematic representation of the underlying state-dependent computational process. (c) Quasi-potential landscape depicting the possible dynamical solutions, that are an intrinsic property of the underlying (signaling/genetic/neuronal) network topology and node dynamics. 
}
\label{fig1}
\end{figure}

When referring to computations, the associated concept usually reflects the formal definition of computation adopted during the first half of the 20$^{th}$ century which was devised with the purpose of answering questions relating e.g. the extent to which mathematics can be reduced to discrete logical formulas, and how mathematical proofs or calculations may be automated \cite{Davis_1978, MacLennan_2004}.
For example, a function on the integer numbers is called \emph{computable}, if an output integer can be calculated by an algorithm after a finite number of steps \cite{MacLennan_2004}. This implies that computation, in an abstract way, refers to a defined mapping between inputs and outputs.
In that broad sense, many basic processes characteristic of living systems on all scales of organization, from single cells in tissues to freely-living single-cell organisms and higher vertebrates, can be defined in terms of computation. 
While vertebrates and many other multi-cellular organisms rely on neuronal networks, single cells and single-cell organisms use protein and/or gene-regulatory networks as computational entities to integrate multi-dimensional sensory information (inputs) with memory, generating complex self-organized behavior (output) \cite{harris1991modulation}. A fox that chases a rabbit, for example, uses its neuronal network to continually processes sensory (visual, auditory, olfactory, tactile etc.) information which is disrupted and changes over time and space, i.e. when the rabbit hides behind a bush (Fig.~\ref{fig1}a, top). In order to avoid random change in the running direction in the absence of a visual contact with the rabbit, the fox integrates the current sensory information with the memory of the last localization of the rabbit to determine its behavior. Immune cells in our body that chase invading bacteria, however, face similar challenges as the fox chasing the rabbit: the cells navigate guided by local chemical cues secreted by the bacteria that are noisy, disrupted, and vary over time and space, in order to engulf and degrade the motile invading microbes (Fig.~\ref{fig1}a, bottom). To avoid immediate switching to random migration when signals are disrupted, single cells, just like the fox chasing a rabbit, require a memory of the localization of the last encountered chemical signal, as a means of generating a reliable migration trajectory over long distances. We will refer to such computations performed by living systems as \textit{biological} or \textit{natural} computations.

This description of processes in living systems through the concept of computation has inevitably led to a frequent referral to terms such as 'circuitry', 'computer/machine', 'execution of programs' or 'interpretation of code'. However, their indiscriminate use can easily convey misleading ideas (and often does) by neglecting fundamental differences between the computational process in living and engineered systems \cite{nicholson_2019, roli_2022, sheetz_2018, dancin_2009, kondepudi_2017,neves_2021, Sol__2022}. For example, computers accept input once at the beginning of the computation (i.e., do not accept signals while performing a particular task), and obtaining the result of the computation requires presentation of the complete input. In contrast, living systems compute in real-time (like the fox and the immune cell), using incomplete inputs which continuously change across time and space. The process of natural computation itself is adaptive, while yielding robust and reproducible responses even in the presence of noise, whereas computer algorithms account only for robustness and reproducibility of computations, which is achieved by minimizing noise (in the constituting computer circuits). Moreover, living systems are characterized by \textit{on-the-fly} and \textit{life-long} learning, features that have not been realized in any man-made system so far.
We therefore discuss in this article the current concepts of computation from a dynamical systems point of view and argue that they explain information processing in living systems only to a very limited extent. We advocate the necessity to base theories of biological computation on transients which can serve as a dynamical basis through which the majority of the observed features of natural computation can be captured\cite{Friston_1997, Mazor_2005, Bondanelli_2020, Maass_2002, Legenstein_2007, Kaneko_2003, Rabinovich_2001, Ashwin_2005_2}. Given that from a conceptual viewpoint, the challenges, which single cells and higher vertebrates face are overlapping to a large degree, we emphasize in this review the necessity for a general theory of natural computations and learning that is applicable to both neural and aneural systems.

\section*{Results}
\section{Computing with stable attractors: machines, neuronal networks and cellular signaling}

The existing theoretical frameworks for biological computations mainly refer to Turing-like computations \cite{turing1939systems}. Let us exemplify this by considering an every-day example of a finite-state machine, a turnstile. A turnstile can be found in two different states, closed ($S_1$) and open ($S_2$). It requires an input, i.e. a coin to switch from the closed to the open state ($S_1 \rightarrow S_2$), and a second input - a push, to return to the closed state ($S_2 \rightarrow S_1$). If a coin is inserted when the system is in $S_2$ (open), or if its bars are pushed when it is in $S_1$ (closed), the machine does not respond (Fig.~\ref{fig1}b). For this system, the computation is realized through the switching between the distinct stable states ($S_1$, $S_2$) that are available, such that the {\textit {specific}} inputs (coin, push) are mapped to a defined state of the turnstile. Despite that this machine has a very limited computational power that does not reflect that of Turing machines in general, we will use it to demonstrate a line of thought that {\textit{(i)}} the state-dependent computations performed by machines display limitations for tasks that are relevant for living systems, and {\textit{(ii)}} range of computations performed by living systems lie outside of the domain of machines, prompting us to propose that broader definition and mechanisms of computations are necessary to describe biological computations.

\begin{figure}[h!]
\includegraphics[width=14cm]{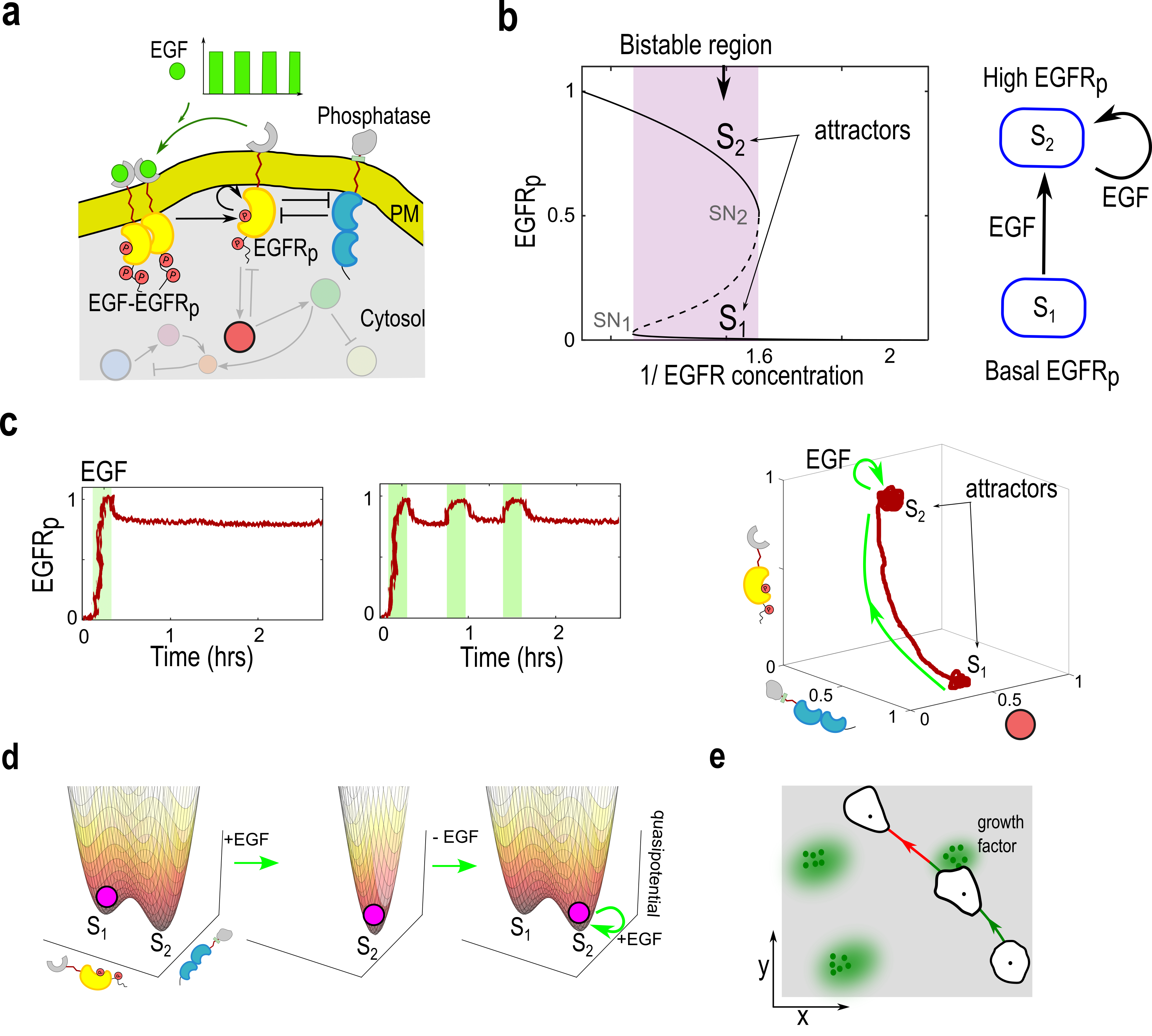}
\caption{Attractor-based computations in cell signaling networks. (a) Cell-surface receptor module enabling sensing of time-varying growth factor signals. EGFRp: phosphorylated epidermal growth factor receptor; EGF-EGFRp: epidermal growth factor (EGF) ligand-bound receptor fraction. (b) Bifurcation diagram corresponding to (a), and a schematic representation of the underlying state-dependent computational process for organization in the bistable regime. SN: sadddle-node bifurcation. (c) EGFRp temporal response to single and multiple EGF pulses for organization in the bistable regime, and a corresponding phase-space trajectory (right). Details in \cite{Stanoev_2018, Stanoev_2020}. (d) Corresponding quasi-potential landscape description. Note the remodeling of the landscape in the presence of transient external signals. (e) Schematic representation of cell migration through disrupted and spatially non-uniform growth factor field for receptor network organized in the bistable regime. The direction is determined by the initial signal the cell encounters. Green/red lines: migration trajectory in presence/absence of signals; arrows denote temporal evolution.}
\label{fig2}
\end{figure}

The computations of the turnstile, in the language of dynamical systems, can be formalized as {\textit {attractor-based computation}}. Turing implicitly used this idea in his unpublished work on intelligent machines \cite{intmach}, as well as in his seminal work on self-organization in living systems \cite{turing1990chemical}. The idea, however, was clearly explicated by Hebb \cite{hebb2005organization} and Hopfield \cite{Hopfield_1982} for neuronal networks, receiving formalization by Hirsch and Baird \cite{hirsch1995computing}: "as the overall system evolves in time, each subsystem passes through a sequence of attractors that are related to specific output of the system, and this sequence is termed as the {\textit{computation process}}". In this view, the number and type of attractors is an intrinsic property of the system, determined by the underlying network topology and nodal dynamics. Thus, restricting only to fixed point dynamics for simplicity, the possible solutions of the system form a so-called quasi-potential landscape which can be explicitly calculated for gradient systems \cite{smale_1961}. Each valley in the landscape corresponds to a stable state, separated by unstable states or saddles (Fig.~\ref{fig1}c). The external signals induce switching between the available states, such that a signal is uniquely associated with a specific valley. This also implies that in absence of a perturbation, the dynamics would be retained indefinitely in a valley.
The same conceptual framework has been also adopted to study signaling networks in single cells. Large number of experimental and theoretical studies over the past two decades have been focused on identifying the underlying protein- or gene-interaction networks or network modules \cite{Choudhary_2010, Emmert_Streib_2014}, relating the possible attractors with the observed phenotypes or responses \cite{alon2019introduction, Ryu_2015, Santos_2007, Jia_2017, Pillai_2022, Rukhlenko_2022,Adelaja_2021}. The question is however, to which extent attractor-based computations can explain the basic features of natural computations, the simplest being real-time processing of non-stationary signals.

\section{Limitations of the current framework: an example from {\textbf{single-}}cell signaling}

Let us consider a bistable system as a minimal case of multistability, i.e. as a system that has multiple, coexisting states/attractors (cf. Box 1). In the quasi-potential landscape analogy, this corresponds to a landscape with two valleys, separated by a saddle. Generally, bistability can emerge via double negative or positive feedback loops, common motifs in gene regulatory \cite{Wang_2014,Papatsenko_2011}, neuronal \cite{Kogo_2020, Kim_2020}, as well as in signaling networks including, e.g., the experimentally identified Epidermal growth factor receptor (EGFR) network\cite{Stanoev_2018} (Fig.~\ref{fig2}a). For parametric organization in the bistable regime (Fig.~\ref{fig2}b), in this case defined by a certain range of EGFR concentrations on the plasma membrane, the system has two available stable states: basal ($S_1$) and high EGFR-phosphorylated ($S_2$) state.
A pulse of epidermal growth factor (EGF) induces a transition from $S_1$ to $S_2$, however, the high EGFRp state will be maintained even after removal of the EGF signal, as $S_2$ is also a stable attractor. This can be seen from the temporal EGFRp profile, as well as the trajectory of the signaling state of the EGFR network (Fig.~\ref{fig2}c). Thus, addition of subsequent EGF pulses will not lead to further state changes in the system, implying that the cell will remain unresponsive to upcoming changes in the environment. Better understanding of these temporal system's responses can be gained from a quasi-potential landscape description: as noted above, the organization in the bistable regime corresponds to a quasi-potential landscape with two wells, corresponding to $S_1$ and $S_2$ (Fig.~\ref{fig2}d). In absence of a signal, the system resides in $S_1$. Upon EGF addition, the quasi-potential landscape remodels to a single well corresponding to $S_2$, resulting in a robust EGFR phosphorylation. When the EGF is removed, however, the landscape resets to its former double-well shape, but the system remains in the $S_2$ well and thereby is unresponsive to upcoming signals (Fig.~\ref{fig1}c), just like the turnstile does not respond to a coin when already in $S_2$.

That such state-dependent computation limits responsiveness to time-varying signals has {\textbf{also}} been experimentally demonstrated for epithelial cells \cite{Stanoev_2018}, implying broad consequences for the mechanistic understanding of how cells navigate in dynamic spatial-temporal chemical fields, e.g. during wound healing or embryogenesis \cite{L_mmermann_2013,Sano_2005}. We and others have shown using numerical simulations, for example, that the cellular migration will be "locked" in the direction of the initial input, leaving the cell unresponsive to any further spatial or temporal signal changes (Fig.\ref{fig2}e), unless the upcoming signals have $\sim$7 fold higher amplitude 
\cite{Nandan_2023, bluttenschon_2022}. Moreover, this attractor-based framework of computation also does not explain how cells can resolve two competing chemoattractant signals \cite{Nandan_2022, Nandan_2023}, as observed for example for leukocyte navigation
\cite{foxman_1999}. 

These findings therefore suggest that the attractor-based framework reaches its limitations for biological computations in single cells: although it provides an explanation for robustness (i.e. maintaining directional memory when the signal is disrupted), it cannot capture the flexibility of natural computations - adaptation to dynamic signals that vary over space and/or time, and thus processing of dynamic signals in real time. Moreover, the attractor-based description also faces an additional, formal problem. As depicted in Fig.\ref{fig2}d, under non-stationary cues, the quasi-potential landscape that underlies the dynamical structure of the system continuously changes
\cite{verd_2013, Stanoev_2018, Stanoev_2020, Nandan_2022, Nandan_2023}. Since the number and positions of steady states are not preserved, the steady states are not formal solutions of the system. It has been therefore argued across different systems including developmental \cite{Manu_2009, Jutras_Dub__2020, Farjami_2021}, signaling \cite{Stanoev_2020, Nandan_2022, Karin_2023} or neuronal systems \cite{Mazor_2005, Rabinovich_2001, timme_2002, neves_2012}, that trajectories or so-called transient dynamics are better suited to describe natural computations.

\section{Computing with transients: chaotic itinerancy and heteroclinic networks}

\begin{figure}
\includegraphics[width=\textwidth]{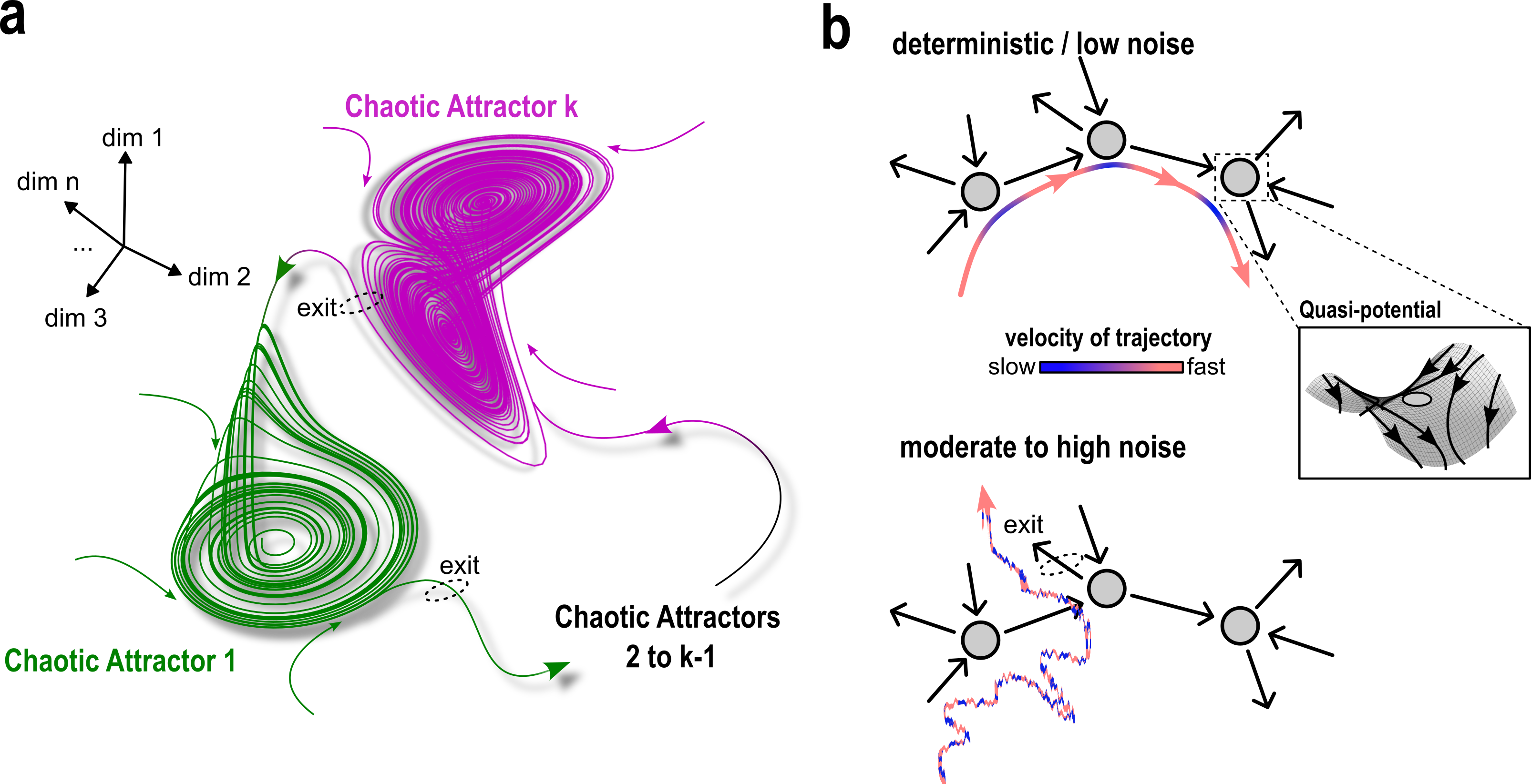}
\caption{Computing with trajectories. (a) Schematic representation of attraction to, and escape from, a quasi-attractor in chaotic itinerancy\cite{Kaneko_2003}. 
(b) Schematic of a heteroclinic channel consisting of multiple saddles (inset) connected such that the unstable manifold of the preceding saddle is connected to the stable manifold of the following in the sequence. Reproducible trajectory guidance can be achieved in the deterministic case or low noise intensities (top), whereas moderate noise levels compromise the reliability (bottom) \cite{koch_2023}. 
}
\label{fig3}
\end{figure}

Trajectories represent the evolution of the state of the system in phase-space. In the presence of attractors, a trajectory will converge asymptotically to a valley in the quasi-potential landscape and remain trapped in it (switching from $S_1$ to $S_2$ in the case of the turnstile and the EGFR system (Figs.~\ref{fig1}b,~\ref{fig2}d)). However, when the landscape continuously changes, the trajectory, thus the dynamics of the system can be maintained away from the attractors \cite{verd_2013, Nandan_2022, Nandan_2023}. In such cases, instabilities have been advocated as an advantageous substitute to attractors, suggesting that systems can exploit unstable states to compute \cite{Ashwin_2005}. One can think of these as instabilities of a special type that have both stable and unstable directions and are referred to as quasi-attractors or attractor-ruins \cite{milnor_1985, Gorban_2004, tsuda_1991}. The presence of the unstable directions allows the trajectory to sporadically switch from one quasi-attractor to another, resulting in the existence of low-dimensional ordered motion. This can be realized for example in chaotic systems, such as networks of nonlinear oscillators (asynchronous neuronal networks) or coupled maps \cite{tsuda_1991, tsuda_1992, kaneko_1990, kaneko_1991}, and is known as chaotic itinerancy (Fig.\ref{fig3}a) \cite{ikeda_1989, tsuda_1991,kaneko_1990}. The transient trapping in the vicinity of the quasi-attractor is proposed to enable robustness via quasi-stability, and since the trajectories do not settle on any attractor, the computations can potentially adapt to dynamic signals. Thus, chaotic itinerancy has been proposed in the general context for natural computation, in particular modeling brain activity, decision making or forming of cognitive functions (cf.\cite{Inoue_2020} and references within). However, this description is limited to systems that display chaotic dynamics.  

Following a growing body of empirical evidence that a large class of natural systems display quasi-stable and sequential (but not chaotic) dynamics, including neuronal firing patterns during olfactory sensing or discrimination tasks \cite{Mazor_2005,Benozzo_2021,Recanatesi_2022}, pattern matching during camouflage in animals\cite{Woo_2023}, cellular signaling systems \cite{Nandan_2022,Karin_2023}, replicator networks \cite{Sardany_s_2007} etc., computation with heteroclinic networks \cite{Mazor_2005, Rabinovich_2001, timme_2002, neves_2012,Ashwin_2005}  has been proposed as a mechanism that generates such sequential dynamics. Heteroclinic objects consist of joined saddles, such that the unstable manifold of the preceding is the stable manifold for the next saddle in the sequence (Fig.~\ref{fig3}b). In such systems, computations are performed when external input signals typically induce long ‘complex’ orbits that pass near a sequence of several ‘simple’ saddles. Moreover, in the course of the temporal evolution, the system's trajectory slows down and is transiently trapped in the vicinity of the saddle fixed point, followed by a quick transition to the upcoming saddle in the sequence (Fig.~\ref{fig3}b, top). However, we have recently demonstrated theoretically that the fidelity of heteroclinic objects to reliably guide trajectories is lost in the presence of moderate noise intensities \cite{koch_2023}. Since the saddle has two unstable directions, noise can easily drive the trajectory away from the path (Fig.~\ref{fig3}b, bottom). This implies that heteroclinic networks are likely not suitable as a generic dynamical framework for computations in living systems, which are inherently noisy.

\section{What can be learned about real-time computations from single cells?}

\begin{figure}[h!]
\includegraphics[width=14cm]{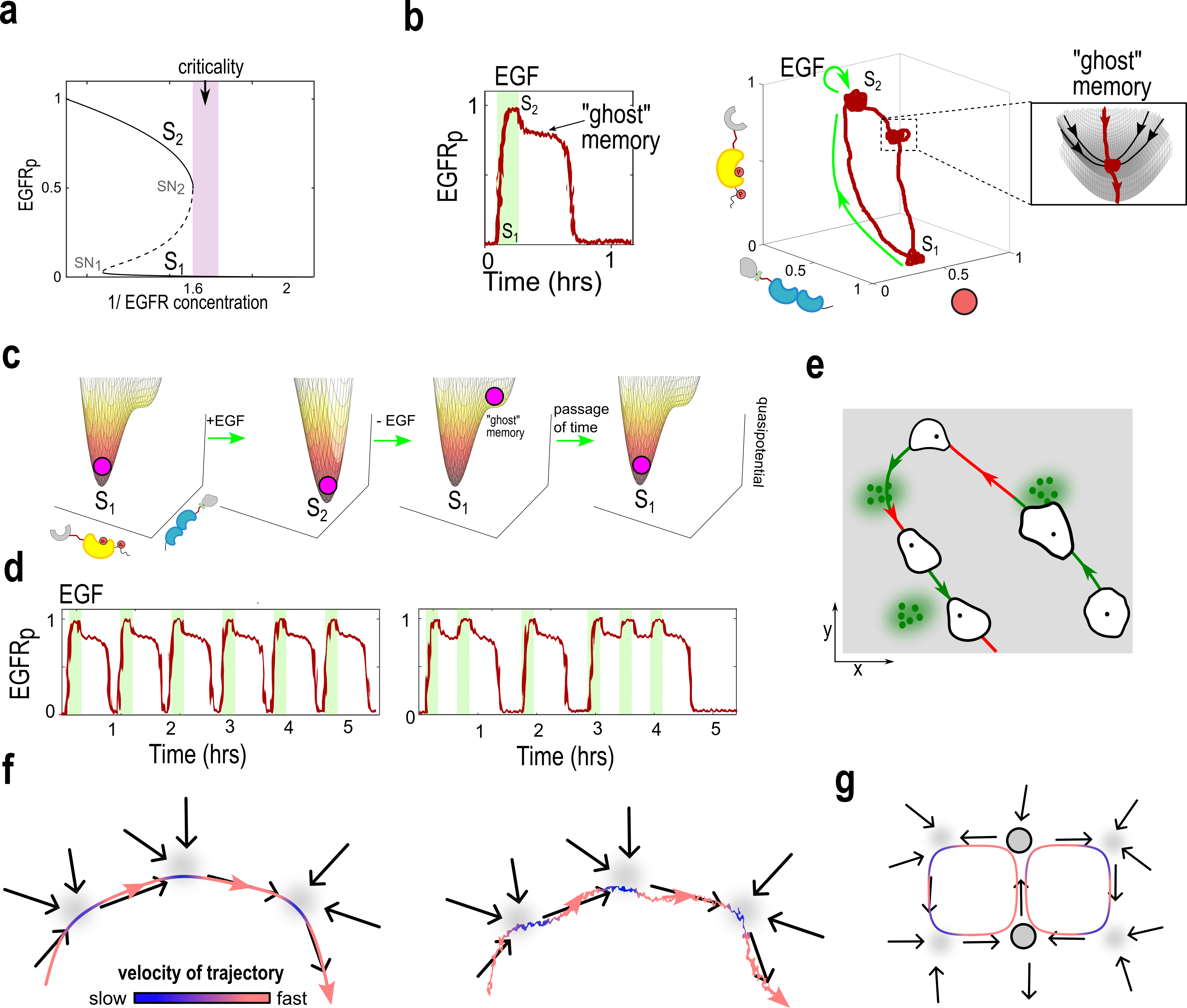}
\caption{Cellular responsiveness to changing signals is enabled for organization at criticality. (a) Bifurcation diagram for the EGFR network depicting organization at criticality. Notations as in Fig.~\ref{fig2}b. (b) Corresponding EGFRp temporal response to single EGF pulse and respective phase-space trajectory. Inset: quasi-potential landscape depicting the geometry of the ghost state. (c) Corresponding quasi-potential landscape changes, showing "ghost" memory state emergence upon signal removal. (d) EGFRp profiles depicting the dependence on the frequency of the EGF signal. (e) Schematic representation of cell migration through disrupted and spatially non-uniform growth factor field for the EGFR network organized at criticality. Cell remains responsive and continuously adapts the migration trajectory to newly encountered signals. Notations as in Fig.~\ref{fig2}e. Details in \cite{Nandan_2022}. (f) Schematic of a ghost channel consisting of multiple ghosts connected such that the unstable direction of the preceding ghost is connected to the stable direction of the following in the sequence. Reproducible trajectories, and thus reliable computing can be achieved even in the presence of noise. (g) Hybrid phase-space object consisted of saddles (circles) and ghost, that can be utilized to generate context-dependent dynamics. See also \cite{koch_2023}.}
\label{fig4}
\end{figure}

As outlined in the introduction, living systems base their behavioral responses on noisy, often conflicting and incomplete time-varying signals, regardless of whether considering a migrating cell following chemotactic cues or a predator hunting its prey. The question therefore arises whether understanding the mechanism of real-time computations for example, for a given single cell task would allow to extract possible dynamical basis on which a generic framework of natural computations can be set.

We recently explored this question by investigating how epithelial cells navigate in dynamic EGF chemo-attractant fields. To establish long-distance migration trajectory, cells would require a dynamic memory to integrate the environmental signals and to quickly adapt the computations following signal changes, while simultaneously allowing for robustness in the responses. Using the EGFR network, we have shown that under physiological conditions, cell do not implement state-dependent computations to solve this task. In contrast, the system is actually organized at \textit{criticality} \cite{Stanoev_2018, joshi_2023}, i.e. exactly at the border between mono- and bistability (Fig.~\ref{fig4}a), where in absence of a growth factor signals, only one steady state, $S_1$, is stable. This organization turns out to be crucial for epithelial cells to sense time-varying signals. What is then here the mechanism of computation?

To understand how cells compute in which direction to migrate when signals change over space and time, we used the quasi-potential landscape description of the EGFR system in conjunction with the system's state trajectories. As mentioned before, the quasi-potential landscape changes under influence of time-varying EGF signals (the simplest being a single EGF pulse). In this case, upon signal addition, the landscape remodels such that the $S_1$ well is lost and the $S_2$ well is stabilized. This landscape remodeling guides the system's state trajectory from $S_1$ to $S_2$, enabling robust EGFR phosphorylation. However, when $S_2$ is lost upon signal removal, on its place a region with a shallow slope is formed. Due to the topological characteristics of this region (three attracting and a single repelling direction, which canalizes the flow (Figs.~\ref{fig4}b)), the trajectory is transiently trapped in a slow region before transiting rapidly to $S_1$. This slow movement of the trajectory is manifested as maintained high EGFRp for a transient period of time after EGF removal (Fig.~\ref{fig4}b,c). Dynamically, this corresponds to a {\textit{ghost}} state \cite{strogatz_1994, Stanoev_2018}. The prolonged EGFRp due to the presence of the ghost therefore acts as a temporal memory that the cell encountered an EGF signal. In single epithelial cells, this memory lasts $\sim$40min on average \cite{Nandan_2022}, which  generates a slow time-scale in the system, despite the sub-minute timescale of phosphorylation-dephosporylation events. 

Cells then use this memory to integrate signals over time. For example, if the following EGF pulse is received while the trajectory is in the ghost, the time the systems stays in the ghost will be prolonged, resulting in longer overall high EGFRp activity. In contrast, if the pulse is received outside of the ghost, the system re-sets to $S_1$ and generates another loop in phase-space, resulting in distinct temporal EGFRp profile (Figs.~\ref{fig4}d). Thus, ghosts aid differential integration of time-varying EGF signals depending on their frequency \cite{Stanoev_2020}. 
Since ghosts enable for information to be stored over time, while still maintaining responsiveness to upcoming signals, they correspond to a mechanism of a 'working memory' on the level of single cells. We have shown both theoretically and experimentally that the working memory of the EGFR network is necessary for epithelial cells to robustly maintain directional migration when signals are disrupted, while remaining sensitive (adaptive) to newly encountering chemoattractant signals (Fig.~\ref{fig4}e) \cite{Nandan_2022, Nandan_2023}. 
From a computational perspective, this example demonstrates that single cells do not rely on attractor-based computations, but rather utilize transient states i.e. ghosts, to maintain information about previously sensed signals, whereas the quasi-stability of these states enables the integration of complex environmental signals to determine a robust, yet adaptable response. Ghosts could therefore potentially serve as an elementary mechanism of natural computations, as they enable to explain real-time response, flexibility, robustness, and adaptability of natural computations.

However, to capture complexity of natural computations, such as sequential and context-dependent responses for example \cite{Kato_2015, Woo_2023}, the underlying biochemical and/or neuronal network likely necessitates composite ghost-based phase-space objects that can give rise to differential dynamics. In a recent theoretical work, we have taken the first steps and, based on a geometrical perspective on ghost states, demonstrated that multiple ghosts can be aligned to form ghost channels (Fig.~\ref{fig4}f) or ghost cycles, in the same way that saddles can form heteroclinic objects (i.e. Fig.~\ref{fig3}b). These structures enable reproducibility of the dynamics, even under large noise intensity, indicating that they are likely suitable model for natural computations \cite{koch_2023}. The hypothesis that ghost scaffold could underline natural computations is further strengthened with the fact that ghost cycles, for example, are typical for gene-regulatory network models \cite{Jutras_Dub__2020, Farjami_2021, Kelsh_2021}. Additionally, it is straightforward to describe how hybrid phase-space objects composed of ghost and saddles for example, can arise in general neuronal or signaling network models (schematic shown in Fig.~\ref{fig4}g). Such structures bear high resemblance to manifolds that emerge from the analysis of neuronal dynamics of simple animals \cite{Kato_2015}. Hybrid ghost scaffolds thereby offer a promising new framework for studying natural computation, as they potentially reproduce dynamical features observed in living systems at different scales.

\section*{Computations at criticality as a possible road forward}

Living systems are no trivial input-output devices, and even single cells or 'simple' species are capable of a surprising variety of complex behaviors, such as memory, learning, hunting, social interactions etc. \cite{Dexter_2019, Gunawardena_2022, Krishnamurthy_2023, Alem_2016,Dussutour_2021}. This property is inevitably related to the nature of biological computations: living system generally balance between opposed features, for example robustness and adaptability. The classical computational frameworks, such as the attractor-based computations that also reflects the functioning of the computers and machines (in an abstract sense), on the other hand, account only for one of these features - the robustness. Thus, a cell that would implement attractor-based computations will remain unresponsive in a dynamic physiological environment, as we have argued using the example of the EGFR network. Computation with unstable states or saddles, on the other hand, does not fulfill the requirement of reproducible transients, such that robustness in the responsiveness cannot be guaranteed. The ghost states in contrast, demonstrate a potential to serve as a basic computational 'unit' of natural computations. These states typically occur for organization at criticality, i.e. at the transition between two distinct dynamical regimes (e.g. between mono- and bistability).

Indeed, criticality has long been a candidate for a governing principle of living systems, especially brain dynamics \cite{Legenstein_2007, Mora_2011, Chialvo_2010, Levina_2007,Seoane_2019}. In this context however, criticality has been discussed as a transition between ordered and a chaotic regime, where the former refers to the (neuronal) activity fading away quickly into a featureless attractor and the latter to a regime where slightly similar initial conditions lead to diverging dynamics. The chaotic regime thus erases any correlation between potentially related inputs and results in trajectories without computational significance. The so-called criticality or 'edge of chaos' separates both behaviours, balancing the stability of the ordered phase and the versatile dynamics of the chaotic dynamics. The chaotic itinerancy we have discussed however, can be also realized via critical states \cite{tsuda_2015}, which implies that organization at criticality could potentially provide a link between different computational 'units' in a dynamical sense that satisfy the conditions of natural computations. Defining a framework of computation with trajectories \cite{Gorban_2004} that relies on a combination of chaotic itinerancy, ghosts and saddles for example, likely has a tremendous potential not only for the description of biological computations in neuronal and signaling networks, but also for understanding the peripheral nervous system, collective responses at the tissue scale or inter-organ communication. 

It becomes evident therefore that there's a need to broaden the existing definitions of computation, which predominantly revolve around machine operations. Specifically, a framework for natural computation must encompass real-time responsiveness, the simultaneous presence of flexibility and robustness in responses, the facilitation of anticipatory processes that preemptively prepare for actions, adaptability in computations, and the enduring capacity for lifelong learning. These attributes stand as fundamental characteristics of living systems, demanding a more comprehensive understanding within the framework of computational theory. To which extent computing with transients can incorporate learning, and in particular on-the-fly and life-long learning, is another exciting open question. Classical frameworks of learning rely on mechanisms that change the network topology (e.g. Hebbian learning generating a 'hard-wired' memory or 'memory-as-attractor' \cite{hebb2005organization}). However, a study aimed at understanding the mechanisms how artificial neuronal networks (ANNs) learn tasks dependent on time-varying inputs, for example a memory-demanding, two-point moving average (of input pulses given at random times) task, showed that the ANN required a set of slow points in a plane (which we interpret as a {\textit{multi-dimensional ghost}}), rather than a plane-attractor \cite{Sussillo_2013}, implying that quasi-stable states likely aid learning. In another example, using the topology of the neuronal network of {\textit{C. elegans}} and single neuron Ca$^{2+}$-dynamics recordings, it has been also recently shown that a consistent behavior in a response to time-dependent inputs is realized via a 'soft-wired' memory  \cite{casal_2020} - a memory that does not require structural changes in the network connectivity, but relies only on the system dynamics for encoding. Such a 'soft-wired' memory could emerge from a ghost state, as we have argued above. These findings therefore suggest that transiently stable states, in addition to being utilized by cells, likely also enable artificial and natural neuronal networks to learn and compute complex tasks. Thus, identifying a mechanism of learning that relies on transiently stable states could potentially open the avenue to study both theoretically and experimentally learning on the level of single cells and single-cell organisms \cite{Dexter_2019,Gershman_2021, Gunawardena_2022, Fitch_2021}. 
We therefore suggest that developing an integrated framework of computations relying on transient quasi-stable dynamics can potentially help to understand how organisms having neuronal networks, but also single cells and cellular communities actively process information and learn from their continuously changing environment to adapt and stabilize their phenotype.

\section*{Box 1: concepts from nonlinear dynamics}
\fbox{\begin{minipage}{40em}
\begin{itemize}
    \item \textbf{Phase-space}: When studying a system of interest (e.g. a cell, neuronal circuit, etc.), a mathematical description captures the evolution of the dynamics of the relevant system variables. These can be for e.g. the concentrations of proteins in a cell, ionic currents in a single neuron, or the instantaneous firing rates of individual neurons in a neuronal network. Together, the space spanned by the system's variables is called the phase space. A point in phase space represents a joint state of all variables in the system.

    \item \textbf{Phase-space flow:} Closely neighboring areas in phase space often exhibit similar dynamics, allowing to understand the system behavior by visualizing the direction in which the system moves from a collection of nearby points via arrows. The resulting vector field indicates the flow in phase space.

\item \textbf{Trajectory in phase-space:} A set of vectors that describes the evolution of the state of the system over time.

\item \textbf{Attractors:} Abstract objects in phase-space towards which the system's trajectories converge. The area or volume in phase-space from which the trajectories are attracted is called basin of attraction, implying that small perturbations only lead to short-lived deviations from the asymptotic dynamics. Different types of attractors include: a stable fixed point (often also called a steady state), stable limit cycles (oscillations) or chaotic attractors. 

\item \textbf{Unstable objects:} Unstable objects include unstable fixed points/limit cycles and saddles. When in the vicinity of an unstable object, the system's trajectories diverge, in contrast to attractors. Saddles are special types of unstable objects that attract the trajectories from two directions along the stable manifold, and repell them from two directions along the unstable manifold. 

\item \textbf{Quasi-potential landscape:} Geometric description of the over-all system's dynamics, where valleys correspond to attractors, separated by saddles or unstable states.

\item \textbf{Bifurcation:} Parameter value for which a sudden qualitative change in the dynamics of a system occurs. A bifurcation diagram is a graphical depiction of the bifurcations in a system using the values of parameters and the variables plotted against each other.

\item \textbf{Criticality:} System's parameter organization at a transition between two dynamical regimes (in the vicinity of a bifurcation), for example between a fixed point and a limit cycle.

\item \textbf{Ghosts:} Phase-space objects characterized with shallow slope in the quasi-potential landscape that funnels system's trajectories towards an unstable direction. 

\item \textbf{Quasi-stability:} Transient stability of the system's dynamics by trapping the trajectory in a specific phase-space area, e.g. due to a ghost state. Note that the time spent in this area can be considerably longer than the time-scales of the underlying processes.

\item \textbf{Noise:} Stochastic, random-like fluctuations in the system dynamics due to processes like Brownian motion, variability in gene expression etc.

\end{itemize}
\end{minipage}}

\newpage

\section*{Acknowledgments}
A.K. acknowledges funding by the Lise Meitner Excellence Programme of the Max Planck Society, D.K. is funded by a fellowship from the European Molecular Biology Organization (Grant nr. ALTF 310-2021).


\bibliographystyle{abbrv}


\end{document}